\documentclass[a4paper,fleqn,usenatbib]{mnras}

\usepackage{newtxtext,newtxmath}

\usepackage[T1]{fontenc}
\usepackage{ae,aecompl}

\usepackage{graphicx}	
\usepackage{amsmath}	
\usepackage{amssymb}	
\usepackage{siunitx}

\newcommand{\emin}{e_{\mathrm{min}}}
\newcommand{\omdot}{\dot{\omega}}

\title[PSR J2051$-$0827 companion's quadrupole]{First measurement of the total gravitational quadrupole moment of a black widow companion}

\author[G. Voisin et al.]{
	Guillaume Voisin$^{1,2}$\thanks{E-mail: guillaume.voisin@manchester.ac.uk;\newline astro.guillaume.voisin@gmail.com},
	C. J. Clark$^{1}$,
	R. P. Breton$^{1}$,
	V. S. Dhillon $^{3,4}$,
	\newauthor
	M. R. Kennedy$^{1}$,
	D. Mata-S\'anchez$^{1}$
	\\
	$^{1}$ Jodrell Bank Centre for Astrophysics, School of Physics and Astronomy, The University of Manchester, Manchester M19 9PL, UK\\
	$^{2}$ LUTH, Observatoire de Paris, PSL Research University, 5 Place Jules Janssen, 92195 Meudon, France\\
	$^{3}$ Department of Physics and Astronomy, University of Sheffield, Sheffield S3 7RH, UK\\
	$^{4}$ Instituto de Astrof\'isica de Canarias (IAC), E-38200 La Laguna, Tenerife, Spain
}
\date{Accepted XXX. Received YYY; in original form ZZZ}

\pubyear{2015}

\begin{document}
\label{firstpage}
\pagerange{\pageref{firstpage}--\pageref{lastpage}}
\maketitle

\begin{abstract}
We present the first measurement of the gravitational quadrupole moment of the companion star of a spider pulsar, namely the black widow PSR J2051$-$0827. To this end we have re-analysed radio timing data using a new model which is able to account for periastron precession caused by tidal and centrifugal deformations of the star as well as by general relativity. The model allows for a time-varying component of the quadrupole moment, thus self-consistently accounting for the ill-understood orbital period variations observed in these systems. Our analysis results in the first detection of orbital precession in a spider system at $\dot{\omega} = -68.6_{-0.5}^{+0.9}$ deg/yr and the most accurate determination of orbital eccentricity for PSR J2051$-$0827 with $e = (4.2 \pm 0.1) \times 10^{-5}$.
We show that the variable quadrupole component is about 100 times smaller  than the average quadrupole moment $\bar{Q} = -2.2_{-1}^{+0.6} \times 10^{41} \si{kg.m^2}$ . We discuss how accurate modelling of high precision optical light curves of the companion star will allow its apsidal motion constant to be derived from our results. 
\end{abstract}



\section{Introduction}

Spider pulsars are binaries composed of a millisecond pulsar primary and a low-mass semi-degenerate secondary orbiting with a sub-day period that is often as short as a few hours. In the event that the companion has a very low mass, $<0.1\, \textrm{M}_{\odot}$, the system is called a black widow. Those with heavier companions ($\sim 0.1$--$0.4$\,M$_{\odot}$) are called redbacks. These binaries are named after two spider species which share the characteristic that the larger and more massive female occasionally eats the smaller male after mating. In their astrophysical counterparts, the companions are slowly evaporated by the pulsar \citep{fruchter_millisecond_1988} after the pulsar has been spun up to millisecond periods by mass transfer \citep{alpar_new_1982}. While it is not entirely clear whether the companions are entirely destroyed \citep{polzin_long-term_2019}, there is evidence that the recycling process in Spiders is particularly efficient, with the two fastest known spinning pulsars being respectively a redback, PSR 1748$-$2446ad \citep{hessels_radio_2006}, and a black widow, PSR J0952$-$0607 \citep{bassa_lofar_2017}. Note that in itself this does not imply that these objects are more massive as spin-up can be achieved with only a small amount of mass being accreted (see \citet{tauris_recycled_2016} and references therein). The observed similarity of the companion with low-mass X-ray binary secondaries as well as the transition of some redback systems \citep{archibald_radio_2009,papitto_swings_2013, bassa_state_2014} to and from accreting states certainly supports the idea of these latter systems being the missing links in millisecond pulsar evolution. There are still a number of mysteries surrounding the evolution of spiders, such as to whether redbacks and black widows are directly related or on two distinct pathways \citep{chen_formation_2013, benvenuto_evolutionary_2012}. Probing the internal state of the companion, through a measurement of its gravitational quadrupole moment, thus represents an invaluable constraint for stellar evolution models.

The high potential of spider pulsars \citep{bochenek_feasibility_2015, roberts_surrounded_2012} for pulsar timing (see e.g. \citet{lyne_pulsar_2012}) is often hindered by their ill-understood orbital period variations, which are usually attributed to fluctuations of their quadrupole moment caused by stellar magnetic cycles as proposed in \citet{applegate_mechanism_1992} and later refined by e.g. \citet{lanza_orbital_1999, lanza_time_2006, volschow_physics_2018,navarrete_magneto-hydrodynamical_2019}. If quadrupole-moment changes are responsible for the observed orbital period variations, then it is possible to design a dynamical model of the binary that includes this contribution. Nonetheless, such variations are only perturbations of a larger quadrupole moment due to the well-known centrifugal and tidal forces. However, the main effect that these latter components generate is an orbital precession which is only detectable if a significant eccentricity is present. This characteristic is made unlikely by the assumed evolution scenario where strong circularising mechanisms are expected (see \citet{voisin_spider_2020} and references therein). 

We recently demonstrated \citep{voisin_spider_2020} that a perturbed orbit cannot be perfectly circular as a direct consequence of Bertrand's theorem which stipulates that only the harmonic and Newtonian potentials can lead to periodic motion. We demonstrated that in the case of spider systems, this property is effectively embodied in an apparent eccentricity and periastron precession. Owing to the estimated magnitude of quadrupole deformation, it was shown that the minimal value of that effective eccentricity may fall well within a detectable range.

In this letter, we report on the application of our model to black widow PSR J2051$-$0827. We re-analysed the timing data published in \citet{shaifullah_21_2016} using an implementation of the model of \citet{voisin_spider_2020}.
%

\section{The timing model}

The timing model presented in \citet{voisin_spider_2020} is an extension of the relativistic binary model of \citet{damour_general_1985, damour_general_1986} to binaries with a companion deformed by tidal and centrifugal forces and accurate to first order in eccentricity. The model comes with the restriction that the spin of the companion is assumed to be synchronized with its orbital motion such that the axis of the deformation remains constant. In addition, a time-varying component is allowed to account for orbital period variations. Thus, binary dynamics derives from the companion's gravitational potential,
\begin{equation}
    \Phi_c =  - \frac{Gm_c}{r}\left(1 + (J_s + J_v(t) + J_t \frac{a^3}{r^3})\frac{a^2}{r^2}\right),
\end{equation}
where $m_c$ is the mass of the companion, $J_s, J_t$ and $J_v(t)$ are dimensionless parameters quantifying the spin, tidal and variable quadrupole components respectively, $G$ is the gravitational constant, $a$ the separation between the pulsar and its companion at an arbitrary reference time $t_0$ and $r$ the distance between the two objects at time $t$ such that $a = r(t_0)$ \citep{voisin_spider_2020}. The quadrupole parameters are related to the quadrupole moment along the radial axis $Q_{rr}$ by 
\begin{equation}
\label{eq:Qrr}
    \frac{3}{2}\frac{Q_{rr}}{m_c a^2} =  J_s + J_v(t) + J_t \frac{a^3}{r^3}.
\end{equation} 
Note that even without the intrinsic variation of $J_v$, equation \eqref{eq:Qrr} shows that the quadrupole moment is subjected to the periodic variations of tidal forces as a consequence of orbital eccentricity. One may therefore define an average quadrupole moment by
\begin{equation}
    \bar{Q} = \frac{2}{3}(J_t + J_s)m_c a^2.
\end{equation}

Relativistic corrections at the first post-Newtonian order are included as in, e.g., \citet{damour_general_1985}, and contribute terms of order 
\begin{equation}
 \epsilon = \frac{G M}{ac^2},   
\end{equation}
 where $M = m_c + m_p$ is the total mass of the system and $m_p$ the mass of the pulsar. 
 
One can show that the tidal and centrifugal components are connected by the relation $J_s =(1+q)J_t/3$, where $q = m_c/m_p$ is the mass ratio of the system. Using equilibrium tide theory (\citealt{sterne_apsidal_1939}; \citealt{ kopal_dynamics_1978}), one can relate these quantities to the apsidal motion constant $k_2$,
\begin{equation}
\label{eq:Jt}
 J_t =  -k_2 \rho_f^5 f^5 q^{-1},
\end{equation}
where $\rho_f = R_f/a$, $R_f$ is the volume-averaged Roche-lobe radius, and $f$ the filling factor of the companion such that $R_c/a = f \rho_f$. It is convenient to use the formula $\rho_f = 0.49q^{2/3} / \left(0.6q^{2/3} + \ln\left(1+q^{1/3}\right)\right)$ \citep{eggleton_aproximations_1983}. The apsidal motion constant depends on integration of the stellar structure, and in particular on the equation of state of the star. 

The variable component is left as a free parameter which is related to the orbital frequency derivatives $f_b^{(i)} = \mathrm{d}^{i}f_b/\mathrm{d}t^{i}(T_a)$ like so,
\begin{equation}
\label{eq:Jv}
	J_v(t) = -\frac{1}{6f_b}\sum_{i = 1} \frac{f_b^{(i)}}{i!}(t-T_a)^{i},
\end{equation}
where $f_b = 1/P_b$ is the orbital frequency at the time of ascending node $T_a$. Interestingly, this translates into $J_v(t) = -\Delta P_b/(6P_b)$ where $\Delta P_b$ is the orbital period variation. 

Together with relativistic effects, quadrupole deformations are responsible for a minimum eccentricity,
\begin{equation}
\label{eq:emin}
	\emin = -J_t(16+q) + \frac{\epsilon}{2}\left(\frac{m_cm_p}{M} + 3\right).
\end{equation}  
The total eccentricity that may be detected in timing observations is the sum of this minimum component and the traditional, hereafter Keplerian, component: $e = \emin + e_K$. The Keplerian component is the one that may be nullified by circularisation processes. On the other hand, $\emin$ should be considered as an effective, as opposed to geometrical, component. In any case, if the eccentricity $e$ is large enough to be detected then one also has to account for orbital precession at an angular rate of
\begin{equation}
\label{eq:omdot}
	\omdot = n_b\left(15 J_t + 3J_s + 3\epsilon\right) 
\end{equation}
where $n_b = 2\pi f_b$. The first term of equation \eqref{eq:omdot} gives the tidal contribution $\dot{\omega}_{\mathrm{tid}}$, the second the spin contribution $\dot{\omega}_{\mathrm{spin}}$ and the third term is the relativistic contribution $\dot{\omega}_{\mathrm{rel}}$.

Since spiders are close binaries, the model only includes the so-called R\oe{}mer delay, namely the geometrical delay induced by the variation of the distance projected along the line of sight of the observer as the pulsar circles its orbit. Relativistic delays such as the effects of time dilation or light bending may safely be neglected \citep{voisin_spider_2020}. It follows that, from pulsar timing alone, the inclination angle of the orbital plane cannot be measured independently from the pulsar semi-major axis, but only the projection of the latter along the line of sight: $x = a_p\sin i$. Additionally, information on the masses of the two stars is limited to the so-called mass function $(m_c\sin i)^{3} /(m_p + m_c)^{2} = G^{-1} x^3 n_b^2$. 

Although one may neglect the relativistic contributions in first approximation, one sees that the mass ratio $q$ remains necessary to derive the apsidal motion constant $k_2$ from equations \eqref{eq:omdot} and \eqref{eq:Jt}. If relativistic corrections are to be included, then the knowledge of both masses is necessary. Furthermore, the filling factor $f$ is certainly the most sensitive parameter needed to derive the apsidal motion constant, as it appears in equation \eqref{eq:Jt} to the fifth power. 
This factor cannot in general be obtained through the technique of pulsar timing, but can instead be extracted from modelling of the optical light curve of the companion (e.g. \citealt{breton_discovery_2013}). The same technique can inform us of the orbital inclination, and thus partially lift the degeneracy of the mass function. The mass ratio can be obtained through optical spectroscopy (e.g. \citealt{van_kerkwijk_evidence_2011}) by measuring the velocity of the companion along the line of sight and comparing it to the projected pulsar velocity derived from timing. If spectroscopic observations are not available, light-curve modelling can provide constraints on the masses of the system, but usually with large uncertainties. As a last resort, one can estimate the range of allowed mass ratios by assuming pulsar masses that lie in the range $1.3 M_\odot \leq m_p \leq 2.4M_\odot$.

\section{Results and discussion}

Here, we re-analysed 21 years of timing data previously published in \citet{shaifullah_21_2016}\footnote{The data processed here is available as online additional material to \citet{shaifullah_21_2016} at \url{http://www.epta.eu.org/aom.html}.} using a version of the ELL1 timing model implemented in the Tempo2 pulsar timing software \citep{hobbs_tempo2_2006, edwards_tempo2_2006} and modified according to the prescriptions of \citet{voisin_spider_2020}\footnote{\label{web:tmodel} The timing model implementation is available here: \url{https://bitbucket.org/astro_guillaume_voisin/spider_timing_model/}}. The uncertainties were estimated using the affine-invariant Markov Chain Monte Carlo (MCMC) algorithm of \citet{goodman_ensemble_2010, foreman-mackey_emcee_2013} with the convergence criterion of \citet{dunkley_fast_2005} (see online material for run details) \footnote{\label{web:MCMC}Our MCMC implementation, with bindings to Tempo2, is available here: \url{https://bitbucket.org/astro_guillaume_voisin/mcmc4tempo2/}}.

\begin{table}
	\renewcommand{\arraystretch}{1.}
	\begin{tabular}{lc}
		Parameter & Value \\
		\hline
		MJD range &  49989.9-56779.3 \\
		NToA & 11391 \\
		Red. $\chi^2$ & $4.06$ \\ 
		RAJ (rad)  &  $5.459064089(37)_{-21}^{+22}$  \\
		DECJ (rad)  &  $-0.147663345(83)_{-79}^{+71}$  \\
		$\mu_\alpha$ (mas/yr)  &  $5.6(24)_{-12}^{+14}$  \\
		$\mu_\delta$ (mas/yr)  &  $3.5(29)_{-48}^{+42}$  \\
		$f$ (s$^{-1}$)  &  $2.21796283653060(66)_{-17}^{+22} \times 10^{2}$  \\
		$\dot{f}$ (s$^{-2}$)  &  $-6.2649(75)_{-12}^{+11} \times 10^{-16}$  \\
		$f_b$ (s${}^{-1}$)  &  $1.167797941(44)_{-25}^{+28} \times 10^{-4}$  \\
		$f_b^{(1)}$ (s${}^{-2}$)  &  $8.(92)_{-15}^{+18} \times 10^{-20}$  \\
		$f_b^{(2)}$ (s${}^{-3}$) &  $-7.(836)_{-100}^{+85} \times 10^{-27}$  \\
		$f_b^{(3)}$ (s${}^{-4}$) &  $-1.0(84)_{-78}^{+66} \times 10^{-34}$  \\
		$f_b^{(4)}$ (s${}^{-5}$)  &  $5.(68)_{-29}^{+35} \times 10^{-42}$  \\
		$f_b^{(5)}$ (s${}^{-6}$)  &  $3.(77)_{-25}^{+30} \times 10^{-49}$  \\
		$f_b^{(6)}$ (s${}^{-7}$)  &  $(7.1)_{-12}^{+9.9} \times 10^{-58}$  \\
		$f_b^{(7)}$ (s${}^{-8}$)  &  $-1.(359)_{-110}^{+91} \times 10^{-63}$  \\
		$f_b^{(8)}$ (s${}^{-9}$)  &  $-2.(79)_{-32}^{+38} \times 10^{-71}$  \\
		$f_b^{(9)}$ (s${}^{-10}$)  &  $4.(25)_{-28}^{+34} \times 10^{-78}$  \\
		$f_b^{(10)}$ (s${}^{-11}$)  &  $1.(37)_{-12}^{+10} \times 10^{-85}$  \\
		$f_b^{(11)}$ (s${}^{-12}$)  &  $-1.0(59)_{-88}^{+73} \times 10^{-92}$  \\
		$f_b^{(12)}$ (s${}^{-13}$)  &  $-4.(91)_{-33}^{+36} \times 10^{-100}$  \\
		$f_b^{(13)}$ (s${}^{-14}$)  &  $1.(73)_{-12}^{+15} \times 10^{-107}$  \\
		$f_b^{(14)}$ (s${}^{-15}$)  &  $1.2(48)_{-79}^{+91} \times 10^{-114}$  \\
		$f_b^{(15)}$ (s${}^{-16}$)  &  $-5.(8)_{-11}^{+11} \times 10^{-123}$  \\
		$f_b^{(16)}$ (s${}^{-17}$)  &  $-1.(69)_{-12}^{+10} \times 10^{-129}$  \\
		$f_b^{(17)}$ (s${}^{-18}$)  &  $-3.(19)_{-23}^{+20} \times 10^{-137}$  \\
		$x$ (lt-s)  &  $4.50705(42)_{-89}^{+86} \times 10^{-2}$  \\
		$\dot{x}$ (lt-s/s)  &  $1.0(75)_{-55}^{+50} \times 10^{-14}$  \\
		$\dot{\omega}$ (deg/yr)  &  $-68.(56)_{-49}^{+91}$  \\
		$T_{\mathrm{asc}}$ (MJD)  &  $5.40910343493(32)_{-58}^{+68} \times 10^{4}$  \\
		$\kappa_s \;(e\sin\omega)$  &  $-8.(2)_{-24}^{+19} \times 10^{-6}$  \\
		$\kappa_c \; (e\cos\omega)$   &  $4.(07)_{-13}^{+11} \times 10^{-5}$  \\
		\hline
		Derived quantities & \\
		\hline
		$e$ & $4.(17)_{-11}^{+11} \times 10^{-5}$ \\
		$\emin$ & $2.2(03)_{-14}^{+14} \times 10^{-5}$\\
		$e_K$ & $1.(96)_{-11}^{+11} \times 10^{-5}$\\
		$J_t$ & $-3.9(08)_{-88}^{+94} \times 10^{-6}$\\
		$J_s$ & $-1.3(26)_{-21}^{+22} \times 10^{-6}$\\			
		$\bar{Q}$ ($\si{kg.m^2}$) & $-2.(16)_{-95}^{+61} \times 10^{41}$ \\ 
		$\epsilon$ & $3.(52)_{-37}^{+35} \times 10^{-6}$\\			
		$\dot{\omega}_{\mathrm{rel}}$ (deg/yr) & $14.3_{-2.0}^{+1.9}$  \\
		$\dot{\omega}_{\mathrm{spin}}$ (deg/yr) & $-5.28_{-0.11}^{+0.12}$  \\
		$\dot{\omega}_{\mathrm{tid}}$ (deg/yr) & $-77.78_{-1.8}^{+1.9}$  
	\end{tabular}
	\caption{\label{tab:results} Results of the MCMC fit of the timing data: mean values of the posterior distribution function are given with their $68\%$ confidence regions. Error bars apply to the digits between parentheses, and to the full number otherwise. The derived quantities other than the eccentricity $e$ have been sampled conservatively assuming a uniform distribution of pulsar masses, $1.3 M_\odot \leq m_p \leq 2.4 M_\odot$, and of $\cos i$ between $0.4$ and $0.8$ corresponding to a mean inclination angle $i\simeq 52\deg$ compatible with \citet{stappers_intrinsic_2001}. }
\end{table}

\begin{table}
\centering
\begin{tabular}{ccccc}
Model & $k$ & $-\chi^2$ & $\Delta \mathrm{AIC}$ & $\Delta \mathrm{BIC}$ \\
\hline
BTX & 27 & 47216 & 1101 & 1079 \\
\textbf{17$f_b$} & \textbf{30} & \textbf{46109} & \textbf{0} & \textbf{0} \\
16$f_b$ & 29 & 46291 & 180 & 172 \\
$f_b^{(6)} = 0$ & 29 & 46113 & 2 & -5 \\
\end{tabular}
\caption{\label{tab:stattest} Comparison between the different models. BTX: model of \citet{shaifullah_21_2016}; 17$f_b$: reference model presented in table \ref{tab:results}; 16$f_b$: reference model with 16 orbital frequency derivatives; $f_b^{(6)} = 0$: reference model with $f_b^{(6)}$ fixed to 0. $k$ is the number of parameters,  $\Delta \mathrm{AIC}$ and $\Delta \mathrm{BIC}$ give the difference of the Akaike Information Criterion and the Bayesian Information Criterion respectively.}
\end{table}

The mean values of the parameter posterior distribution functions are given in table \ref{tab:results}. Compared to the fit using Tempo2's BTX model presented in \citet{shaifullah_21_2016}, the reduced $\chi^2$ is significantly improved (4.06 vs 4.16, corresponding to $\Delta\chi^2 \simeq -1107$, see also table \ref{tab:stattest}) and our model is unambiguously favoured by both the Akaike information criterion (AIC, \citet{akaike_new_1974}) and the Bayesian information criterion (BIC, \citet{schwarz_estimating_1978}): $\Delta \mathrm{AIC} \simeq -1101$ and $\Delta \mathrm{BIC} \simeq -1079$ (see also textbooks such as \citet{burnham_model_2002}).
Some of the parameters common to both fits are significantly different, the largest discrepancy occurring in the spin frequency  $f$  which is different by $\sim 57$ standard deviations. This indicates that when precession and eccentricity are not included $f$ adjusts to partly compensate. The orbital parameters $x, \dot{x}, T_{\mathrm{asc}}, f_b$ 
are consistent within $2\sigma$. The orbital period derivatives show similar but nonetheless significantly different values between the two fits (see below). This might be due to the large correlations between each of these parameters (see the corner plot in the online material). The $1\sigma$ error bars themselves are quite different between the BTX fit and the present one, which is partly the result of the different methods used: Tempo2 returns an estimate based on its least-square fit while here we use an MCMC sampling of the posterior distribution function. We note that $f_b^{(6)}$ is consistent with zero within error bars while the Tempo2 fit returns a narrower uncertainty, inconsistent with zero. In order to ascertain that the number of orbital frequency derivatives required with our model is still 17, we have run another MCMC with 16 derivatives only (from $f_b^{1}$ to $f_b^{16}$). Following \citet{shaifullah_21_2016} we have compared the two runs using AIC but also BIC which showed that 17 derivatives were still favoured (see table \ref{tab:stattest}). This is consistent with the fact that eccentricity and precession do not correlate with orbital frequency derivatives (see supplementary online material) and that all the derivatives of order higher than 6 are significantly measured to finite values. We have also run another MCMC with $f_b^{(6)}$ fixed to 0 and found that, as expected, this model is approximately equivalent to the full model presented in table \ref{tab:results}, only being slightly disfavoured by AIC and favoured, although not strongly, by BIC due to the fact that BIC has a stronger penalty for additional parameters. However, the frequency derivatives are only an empirical modelling that strongly depends on a specific observation span. Thus, the fact that of $f_b^{(6)} \sim 0$ would appear to be a coincidence for this particular time span and reference epoch. In order to compare with \citet{shaifullah_21_2016} we chose to report the 17-derivative model in table \ref{tab:results}. Moreover, we note that the large reduced $\chi^2$ we obtain is either due to un-modelled effects or to underestimated uncertainties on the times of arrival. In the latter case, the error bars given in table \ref{tab:results} should be multiplied by $\sim 2$, particularly since the marginalised posterior distributions of all parameters are closely Gaussian (see online material). 

The true novelty of the present fit is the detection of a large orbital precession together with the most accurate detection of orbital eccentricity (see table \ref{tab:results}). The value of the measured eccentricity is somewhat smaller than was hinted at in previous works that gave upper limits \citep{stappers_orbital_1998, doroshenko_orbital_2001} or firm detections\citep{lazaridis_evidence_2011,shaifullah_21_2016}. However, these detections were made on shorter datasets, and in particular the most significant 1000-day aeon in \citet{shaifullah_21_2016}. On the full 21-year dataset (used both in this work and in \citet{shaifullah_21_2016} the Tempo2 best fit with precession set to 0 returns a smaller eccentricity of $9\pm 4\times 10^{-6}$ detected only at a 2-sigma level. This is consistent with the idea raised in \citet{voisin_spider_2020} that an unaccounted large precession averages out the eccentricity over the time scale of a precession period ($\sim 5$ years in the present case), in the sense that the envelope of a precessing eccentric orbit is a circle. This idea is reinforced by the finding of \citet{shaifullah_21_2016} of an apparently variable eccentricity vector when fitting independently small subsets of times of arrival. Further, our assumption that precession is primarily caused by a large gravitational quadrupole moment of the companion star is supported by the negative sign of the precession rate. Indeed, the other source of precession in the model, namely relativistic precession, can only contribute a positive term to the total rate (see also table \ref{tab:results}).

A knowledge of the mass ratio $q$ and the inclination of the system $\sin i$ is needed to derive the quadrupole parameters $J_s$ and $J_t$. In the case of PSR J2051$-$0827 the inclination was estimated from optical observations of the companion in \citet{stappers_intrinsic_2001}, although with important uncertainties since  $36\si{deg} \lesssim i \lesssim 58\si{deg}$ at $1\sigma$, while the mass ratio has to be estimated from a prescription on the pulsar mass which we take in the range $1.3 \leq M_p \leq 2.4$. Fortunately, the fact that quadrupole-induced precession here dominates over relativistic precession ($16 J_t \gg 3\epsilon$) and that the mass ratio is very small, $ q\ll 1$, render the derivations of the quadrupole parameters little dependent on the values and uncertainties of $q$ and $\sin i$. In other words, to first order one has $J_t \simeq \dot{\omega} / (16n_b)$. To go further, we can use conservative estimates of the pulsar mass and system inclination (\citet{stappers_intrinsic_2001}, see table \ref{tab:results}). In particular, this explains the large uncertainty on the average quadrupole moment $\bar{Q}$ as it depends heavily on both inclination and mass ratio via the factor $m_c a^2$. Nonetheless, the magnitude we derive in table \ref{tab:results} is $\sim 6$ times larger than the value theoretically estimated in \citet{lazaridis_evidence_2011}. This discrepancy can be at least partially explained by the fact that the model used by these authors \citep{lanza_orbital_1999} only includes the spin-induced deformation but neglects the dominant tidally-induced component of the quadrupole momentum. 

In the same manner we determine the minimum eccentricity $\emin$, equation \eqref{eq:emin}, and the Keplerian eccentricity $e_K$. It is interesting to note that the orbit is not perfectly circularised since $e_K$ accounts for nearly half of the total eccentricity. This can be used to derive an eccentricity age \citep{voisin_spider_2020}, $\tau_e = \tau_c \log_{10}(1/e_K)$, where $\tau_c$ is the circularisation time-scale \citep{rasio_tidal_1996, zahn_reprint_1977} which depends on the component masses and the companion surface temperature. The eccentricity age gives an upper bound on the time needed for the orbit to circularise assuming only tidal forces are at work and an initial eccentricity of 1. For PSR J2051$-$0827 we calculate the range $\tau_e = 1.6 - 4.1 \times 10^{8}$ years, where the uncertainty comes primarily from the masses as well as from the night-side temperature of the companion which we take to be $T_c = 2600 - 3200$K \citep{stappers_intrinsic_2001}. Unless another effect drives up eccentricity in this system, one should therefore conclude that the system has been in the black widow state for at most a few $10^8$ years, and likely much less than that.

\begin{figure}
	\centering
	\includegraphics[width=0.9\columnwidth]{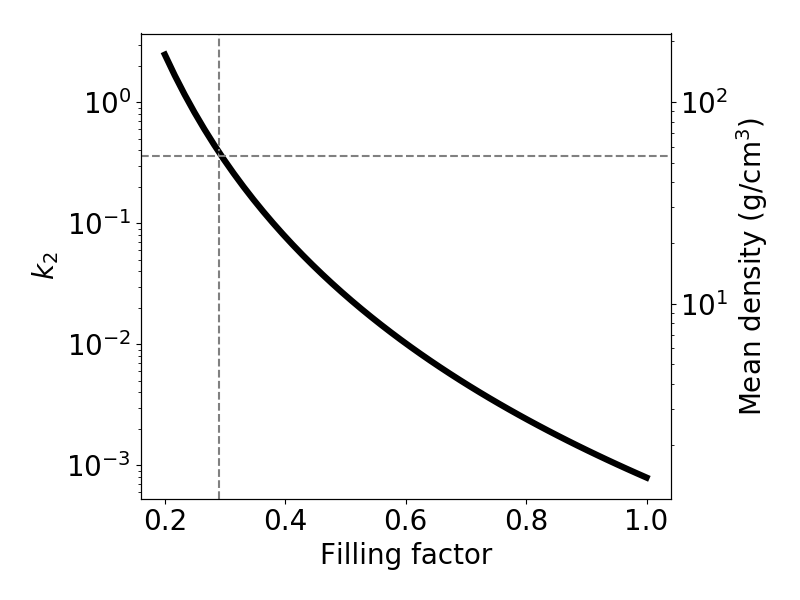}
	\caption{\label{fig:ksvsf} Apsidal motion constant as a function of the filling factor based on the timing results and equation \eqref{eq:Jt}. The thickness of the line includes mass ratios from $q = 0.016$ to $q = 0.022$ and is much larger than uncertainties due to parameters derived from timing. The right vertical axis shows the corresponding mean density. The horizontal dashed line shows the apsidal motion constant PSR J2051$-$0827 would have, $\sim 0.34$, if it had the same mean density of $54$\,g/cm${}^3$ as the densest known black widow companion, PSR J0636+5128 \citep{kaplan_dense_2018}. The vertical dashed line shows the corresponding filling factor which is $\sim 0.29$. }
\end{figure}

Of particular interest is the estimate of the apsidal motion constant $k_2$ since it can be directly related to the internal structure using stellar models. Unfortunately, our knowledge of the filling factor of the companion star, to which $k_2$ is extremely sensitive (see equation \eqref{eq:Jt}), is very poor since it ranges from $0.2$ to $1$. Indeed, \citet{stappers_intrinsic_2001} found that due to the asymmetry of the light curve two solutions were possible: one fitting well when only one-half of the light curve was considered (reduced $\chi^2 = 0.96$) yielding a filling factor $f = 0.43_{-0.16}^{+0.23}$, and another fitting (albeit poorly) the full light curve (reduced $\chi^{2} = 5.6$) with a filling factor of $f = 0.95_{-0.02}^{+0.05}$. Therefore, we show in Figure \ref{fig:ksvsf} the value of $k_2$ as a function of the filling factor. We find that a broad range of stellar structures are possible, since $10^{-3} \leq k_2 \leq 0.3$. As a reference, the Sun has an intermediate $k_2 \simeq 0.015$, while secondary stars in cataclysmic variable systems and hot Jupiters mark the low and upper ends with $k_2 \sim 10^{-3}$ and $0.2$, respectively \citep{ogilvie_tidal_2014, warner_apsidal_1978, kramm_constraining_2012}.\footnote{Our definition of $k_2$ differs by a factor of 2 compared with the value given in \citet{kramm_constraining_2012}. See also \citet{voisin_spider_2020}.}. The uncertainty on the nature of PSR J2051$-$0827's companion was pointed out in \citet{lazaridis_evidence_2011} who argued that it could either be a white dwarf, a brown-dwarf like star or a semidegenerate helium star. We also note that black-widow companions are potentially most similar to hot Jupiters regarding mass and surface temperature. This case would require a small filling factor, typically $f \lesssim 0.5$, compatible with one of the aforementioned light-curve solutions, while a Roche-lobe filling solution, common to spider companions and more similar to cataclysmic variables, would correspond to a particularly small $k_2$. By considering as an upper limit the mean density of the densest known black-widow companion, PSR J0636+5128's \citep{kaplan_dense_2018}, we obtain a lower-limit filling factor of $f \sim 0.29$ giving the upper limit $k_2 \sim 0.34$ (see figure \ref{fig:ksvsf}).
However, figure 4 of \citet{kaplan_dense_2018} shows that the mean densities of black widow companions are scattered from $\sim1$\,g/cm${}^3$ to $\sim54$\,g/cm${}^3$, preventing a more accurate estimate with this criterion. It should also be noted that PSR J0636+5128 may result from a peculiar evolution and not be representative of black widows.

\begin{figure}
	\centering
	\includegraphics[width=\columnwidth]{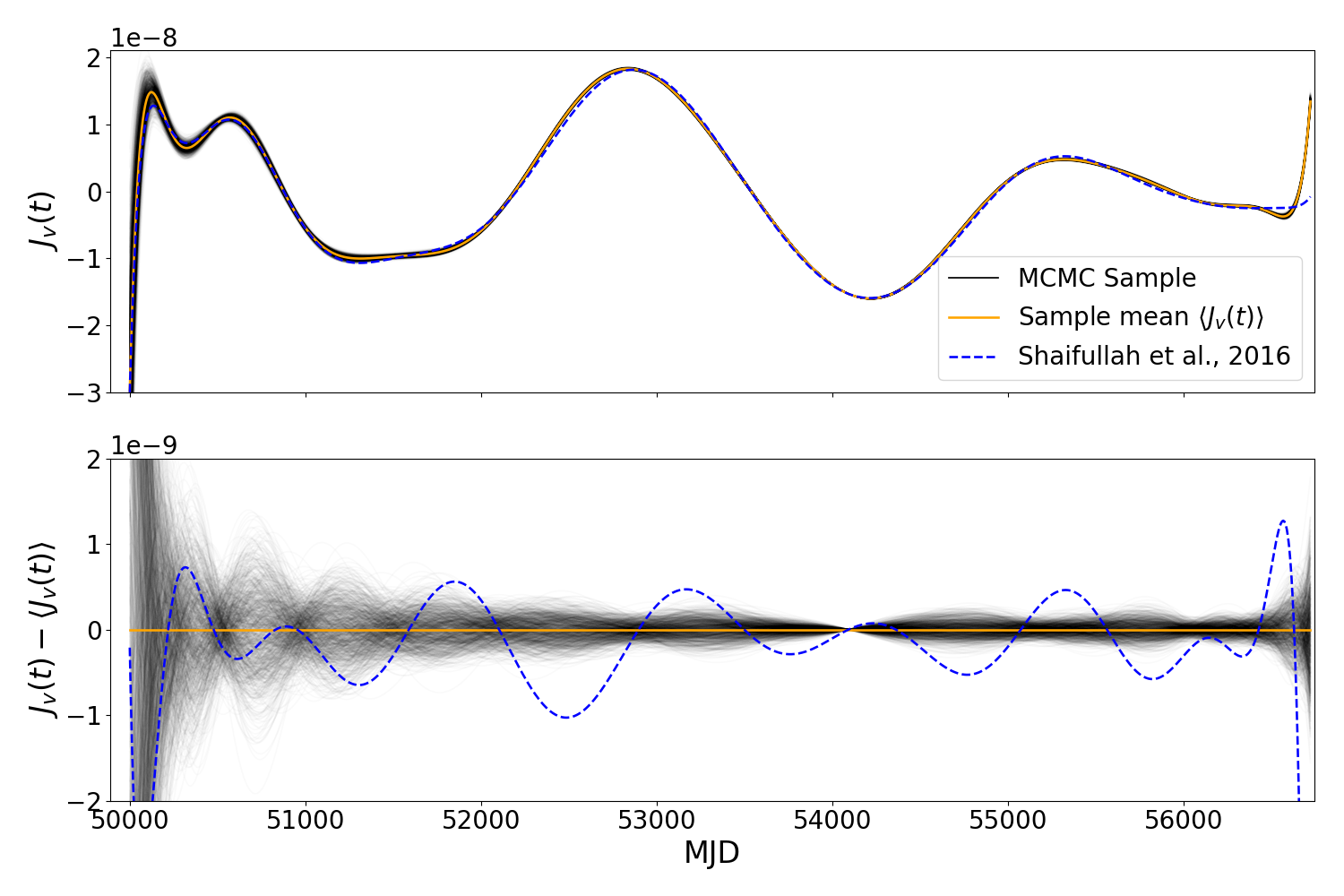}
	\caption{\label{fig:Jv} Upper panel: Variable quadrupole parameter $J_v(t)$ calculated using equation \eqref{eq:Jv}. The mean solution corresponding to table \ref{tab:results} is shown in solid orange, while 1000 lines drawn from the MCMC sample are shown in shades of black. For comparison, the solution of \citet{shaifullah_21_2016} is shown in dashed blue. Lower panel: difference from the upper panel curves and the mean solution, with the same colour code. }
\end{figure}

We show in figure \ref{fig:Jv} the evolution of the variable quadrupole component according to equation \eqref{eq:Jv} and compare it to the curve obtained using the BTX fit of \citet{shaifullah_21_2016}. Although broadly consistent, the two versions show significant differences which cannot be easily reconciled with the inclusion of new parameters since no correlations are expected between eccentricity, orbital precession and orbital period derivatives \citep{voisin_spider_2020}. This is confirmed by our MCMC fit (see online material). The amplitude of the variations, defined as $\|J_v\| = \max(J_v) - \min(J_v) $, are $\|J_v\| \simeq 4.8\times 10^{-8}$ and down to $\|J_v\| \simeq 3.4\times 10^{-8}$ if one neglected the edges of the time series which might be subjected to boundary effects. To ensure that the apparent divergence of the polynomial fit in figure \ref{fig:Jv} is not intrinsic to the data but due to boundary effects we have run our MCMC with the first and last 200 days of data removed and observed again the exact same type of unstability at the edges. As a consequence, the variable quadrupole component is nearly 100 times smaller than the tidally-induced component, consistent with the estimate of \citet{lazaridis_evidence_2011}.

\section{Conclusions}
We have applied a new timing model accounting for the effects of centrifugally and tidally induced quadrupole deformations of the companion star to the black widow pulsar PSR J2051$-$0827, re-analysing the data published in \citet{shaifullah_21_2016}. This led to the first detection of orbital precession in a spider system, $\dot{\omega} = -68.6_{-0.5}^{+0.9}$ deg/yr, which results from the combined effects of an average quadrupole momentum $\bar{Q} =  -2.2_{-1}^{+0.6} \times 10^{41} \si{kg.m^2}$ and of relativistic precession. This also permitted the most precise detection of orbital eccentricity in PSR J2051$-$0827 with $e = (4.2 \pm 0.1) \times 10^{-5}$, which is also the first unambiguous detection of eccentricity using the complete 21-year timing data set without splitting it into [shorter] ``aeons'' \citep{shaifullah_21_2016}. Indeed, over 21 years smearing due to precession averages out the apparent eccentricity to nearly zero if not accounted for.  In addition, the model accounts for orbital period variations by including a time-dependent quadrupole contribution. We could then deduce that this variable component is about 100 times smaller than the combination of tidal and centrifugal deformations. We show that these results can be used to derive the apsidal motion constant of the companion star (figure \ref{fig:ksvsf}) and thus open a new window on the internal structure of these exotic objects. However this will require high-quality optical light curves in order to determine the Roche-Lobe filling factor of the companion to within a few-percent uncertainty, which is within reach of current instruments.

\section*{Acknowledgements}

The authors acknowledge support of the European Research Council, under the European Union's Horizon 2020 research and innovation program (grant agreement No. 715051; Spiders).

The authors acknowledge the use of data from the European
Pulsar Timing Array (EPTA: \url{http://www.epta.eu.org}). The EPTA is
a collaboration of European institutes working towards the direct
detection of low-frequency gravitational waves and the implementation of the Large European Array for Pulsars (LEAP).

The authors would like to thank the anonymous referee the comments and suggestions of whom helped improving significantly the initial draft.

%




\bibliographystyle{mnras}
\bibliography{J2051quadrupole} 



%


\bsp	
\label{lastpage}
\end{document}